\documentclass[journal]{IEEEtran}
\usepackage{amsmath,amsfonts,amssymb,amsthm}
\usepackage[utf8]{inputenc}
\usepackage{graphicx}
\usepackage{cite}
\usepackage{times}
\usepackage{url}
\usepackage[hidelinks]{hyperref}
\usepackage{array,tabularx}
\usepackage{booktabs,multirow}
\usepackage{siunitx}
\newcommand{\TSETabSetup}{%
  \setlength{\tabcolsep}{4pt}%
  \renewcommand{\arraystretch}{1.15}%
}
\newcolumntype{Y}{S[table-format=1.2]}
\newcolumntype{Z}{S[table-format=1.3]}
\usepackage{threeparttable}
\usepackage{colortbl}
\definecolor{lightgray}{gray}{0.8}
\usepackage{pifont}
\usepackage[caption=false,font=normalsize,labelfont=sf,textfont=sf]{subfig}
\usepackage{stfloats,textcomp,verbatim}
\usepackage{algorithm}
\usepackage{algorithmic}
\usepackage{caption}
\usepackage{pgfplots}
\colorlet{wx}{blue}
\pgfplotsset{compat=1.17}
\usepgfplotslibrary{groupplots}
\pgfplotsset{
  every axis/.append style={font=\scriptsize},
  every axis title/.style={font=\scriptsize\bfseries},
  every axis label/.style={font=\scriptsize},
  legend style={font=\tiny, draw=none, fill=white, fill opacity=0.8, text opacity=1},
}
\begin{document}

\title{Exploring a Single Autoregressive LLM for Unified Target Speech
Extraction across Synchronous and Asynchronous Cues}


  \author{Wenxuan Wu, Shuhan Zhang, Shuai Wang,~\IEEEmembership{Senior Member,~IEEE}, Haizhou Li,~\IEEEmembership{Fellow,~IEEE}%
}

\maketitle

\begin{abstract}
Target speech extraction (TSE) typically trains a separate
extractor per cue, and visual-cue systems often need corruption-matched
training to remain robust under visual frame corruption. We show that one
autoregressive LLM backbone, TSE-Omni, can serve both temporally synchronous
cues (lip movements, co-speech gestures) and asynchronous cues (enrollment
audio, text). TSE-Omni is driven by next-token prediction: each step predicts
target speech semantic tokens from its own past outputs, which we term
self-enrollment, forming a continuous target-speech context initialized by the
enrollment cue (asynchronous audio or text, or a short visual prefix). This
enables audio-visual compensation: the model uses synchronized visuals when
intact and its token history when visual frames are missing. Under clean
visuals, TSE-Omni matches strong discriminative and generative baselines
(SpeechBERTScore 0.81 on VoxCeleb2 and 0.89 on LRS3 zero-shot) with higher
DNSMOS. On the same VoxCeleb2 test set, after a 2\,s clean visual start,
removing the remaining visual frames leaves SpeechBERTScore at 0.81. It remains
usable under sparse overlap and multi-speaker interference, and supports
streaming inference. Project page:
\href{https://alexwxwu.github.io/tseomni-main/}{https://alexwxwu.github.io/tseomni-main/}.
\end{abstract}


\section{Introduction}

\begin{figure}[h]

\includegraphics[scale=0.23]{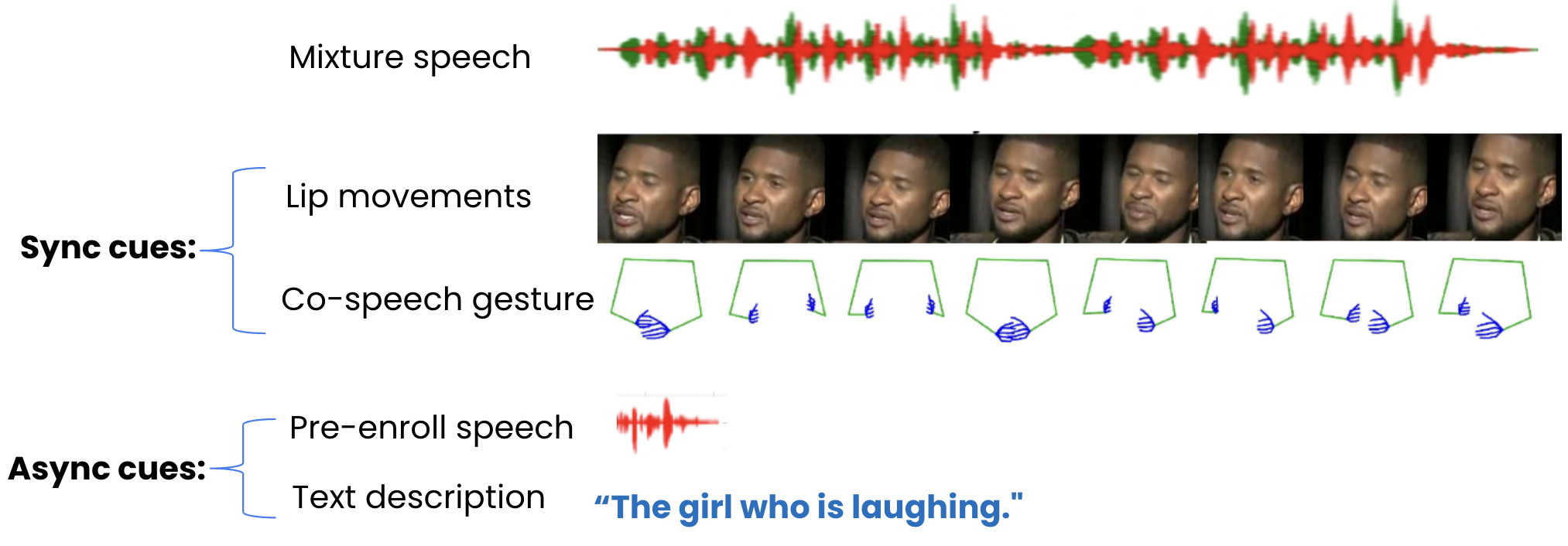}
\caption{Cue types. Cues are either \emph{temporally synchronous} with the
mixture (lip movements, co-speech gestures) or \emph{temporally asynchronous}
(pre-recorded audio, text).}
\label{fig:cue_type}
\end{figure}

Target speech extraction (TSE) aims to extract the speech of a target
speaker from a mixture of multiple talkers, based on a cue that indicates
\emph{who} to extract, as illustrated in Fig.~\ref{fig:cue_type}. Cues
fall into two categories. \emph{Temporally synchronous} cues, such as lip
movements and co-speech gestures, are aligned frame-by-frame with the
mixture and can track the target speaker over time. \emph{Temporally
asynchronous} cues, such as a pre-recorded audio clip or a text
description, are not time-aligned; they help identify the target but
cannot track it continuously.

Audio- and visual-cue-based methods are the most widely studied
paradigms. Audio-based approaches, such as SpEx~\cite{spex}, usually guide
separation with a short pre-recorded enrollment audio, whereas visual-based
methods, such as TDSE~\cite{tdse}, rely on synchronous lip movements. Recent
work provides richer context: USEF~\cite{zeng2025usef} and
SoloSpeech~\cite{solospeech} apply cross-attention between the audio cue and
the mixture speech; AVHuMAR~\cite{ijcnn} and $C^2$AVTSE~\cite{wu2025c}
integrate audio-visual context with a mask-and-recovery strategy; and
MeMo~\cite{li2025memo} builds memory banks from historical speech context
for online extraction. Despite this progress, most TSE systems are still
optimized mainly at the signal level, without leveraging the phonetic and
linguistic knowledge that speech carries.

Meanwhile, large language models (LLMs) have been introduced to TSE.
ELEGANCE~\cite{wu2025incorporating,wu2026elegance} injects textual
linguistic guidance into TSE models. This textual alignment improves the
semantic consistency between the extracted speech and the ground truth,
and also helps compensate for corrupted visual cues at no extra inference
cost. However, its extraction is still not purely semantic-driven.

\begin{figure}[h]
\centering
\includegraphics[scale=0.35]{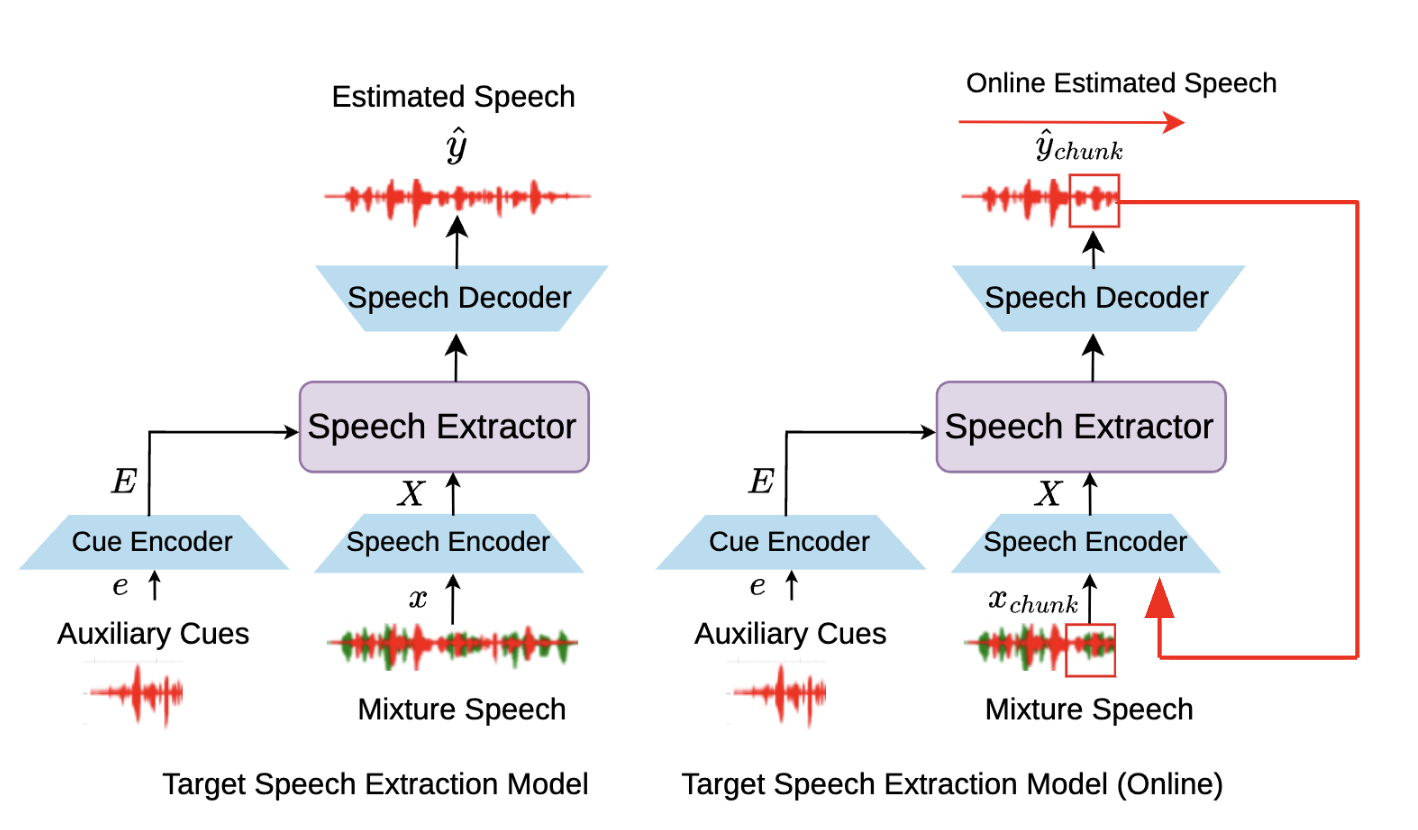}
\caption{Self-enrollment compensation: historical predicted target-speech semantic
tokens are fed back as a running enrollment cue. This self-loop is inherent to
AR decoding.  TSE-Omni
is the first to exploit it to compensate for corrupted visual cues without a corruption-matched training stage in AR LLM framework.}
\label{fig:self_enroll}
\end{figure}

Some studies use LLMs as speech extractors. LauraTSE~\cite{tang2025lauratse},
built on LauraGPT~\cite{du2023lauragpt}, predicts the first two RVQ layers
conditioned on the audio cue and the mixture embeddings in an AR stage,
then predicts the remaining codebooks in a NAR stage. LLaSE-G1~\cite{kang-etal-2025-llase}
unifies speech and audio enhancement with a single-layer codec. These
models move from signal-level processing toward semantic-driven
extraction, but disentangling the target speech from the mixture remains
their bottleneck and the main source of content and acoustic
hallucination. Extending them to multiple cues under limited parameters and
training data is therefore difficult. Their autoregressive formulation,
however, offers a natural remedy: during decoding, an AR model accumulates
a history of its own predicted target-speech tokens, which can be reused
as a continuously refreshed enrollment cue at no extra cost. We term this
mechanism \emph{self-enrollment} (Fig.~\ref{fig:self_enroll}); it
supplies speaker identity and linguistic context that persist even when an
external cue degrades.

However, existing AR-LLM TSE systems remain confined to a single cue:
they accept either an audio cue or a visual cue, but not both, and are
trained as separate extractors, so the AR token history is never reused to
compensate a missing visual stream. Such single-cue systems fail in two common
scenarios. When later visual frames are corrupted or missing (e.g., the target
moves out of view), a purely visual model has no fallback and often needs
corruption-matched training~\cite{wu2024target_cvpr} or a dedicated recovery module~\cite{ImagineNET}. When the target
switches mid-mixture, an audio-only model must condition on two audio cues
at once, which is ambiguous, and errors made at the switch point
accumulate over time.

To address both failure modes and unify different cues within a single model, this study proposes TSE-Omni, one AR-LLM backbone that handles both synchronous and asynchronous cues. It predicts target speech tokens from its own history (self-enrollment), fusing them with synchronized visual tokens when available, so it can compensate for corrupted or missing visual frames without corruption-matched training, and can follow a target switch mid-mixture. Our main contributions are as follows:

\begin{itemize}
\item We propose TSE-Omni, which supports temporally synchronous and asynchronous cues with one AR-LLM extractor backbone, rather than a separately trained extractor per cue.
    \item We introduce self-enrollment (Fig.~\ref{fig:self_enroll}): the model conditions on its own accumulated speech tokens, so zero-filled corrupt or missing frames (a case seen in omni training) are compensated from the speech-token history, with no corruption-matched training and no recovery module.

    \item Experiments on VoxCeleb2, plus zero-shot LRS3, visual corruption, visual-guided target switching, sparse overlap, multi-interferer mixtures, and streaming, show that TSE-Omni matches strong baselines on clean-visual semantics while improving perceptual quality, and that SpeechBERTScore is unchanged on the core test set when later visual frames are zero-filled.
\end{itemize}

\section{Related Work}

\subsection{Audio-Based and Visual-Based TSE Models}

\subsubsection{Audio-Based TSE Models}
These models primarily depend on the target speaker's voice characteristics.
Typically, a short pre-recorded speech clip (a few seconds in duration) is
used as the enrollment cue. A speaker encoder extracts a speaker embedding
from this clip, which subsequently guides extraction of the target speech from
the mixture signal, as in SpEx~\cite{spex}. More recent studies go beyond
speaker identity and incorporate broader contextual information from the cue
speech. For example, instead of relying solely on speaker encoders,
USEF~\cite{zeng2025usef} and SoloSpeech~\cite{solospeech} employ a
cross-attention mechanism between the cue and mixture speech to capture
contextual information and reduce latent feature mismatch.

\subsubsection{Visual-Based TSE Models}
These models leverage temporally synchronized visual cues, such as lip
movements. Typically, a visual speech recognition (VSR) front-end is employed
to extract visemes from the target speaker's lip movements, which are then
aligned with the target speaker's phonemes in the mixture speech, as in
TDSE~\cite{tdse}. However, visual corruption frequently occurs in practice and
degrades performance. To mitigate this, methods such as
ImagineNET~\cite{ImagineNET} adopt an interleaved extractor, where intermediate
speech refines corrupted visual cues, and the refined visual cues in turn
enhance speech extraction. This iterative design proves effective under
severe visual degradation.

Existing audio- and visual-based extractors are developed and deployed independently,
so a separate backbone is stored for each cue. This increases deployment cost and
makes it harder to reuse complementary cross-modal information at inference.
TSE-Omni instead shares one AR-LLM extractor across audio- and visual-cue TSE,
with a modality-specific front-end for each cue.

\subsection{Target Speech Extraction with LLMs}
Large language models (LLMs) have shown strong capabilities in speech
enhancement, separation, and extraction. Current LLM-based TSE methods follow
two main paradigms.

The first approach uses LLMs as knowledge bases to provide linguistic cues. For
instance, studies such as~\cite{wu2025incorporating,wu2026elegance} extract
textual linguistic guidance, including linguistic priors, constraints, and
predictions, from models like RoBERTa~\cite{Liu2019RoBERTaAR} and
Qwen3~\cite{qwen3technicalreport}, and inject this information into traditional
TSE pipelines. This textual guidance helps compensate for corrupted visual cues,
thereby improving semantic coherence and speech quality.

The second approach fine-tunes LLMs directly for speech enhancement or TSE
tasks~\cite{gense,Genhancer}. For example, GenSE~\cite{gense} employs an AR-LLM
to predict clean speech semantic tokens from noisy inputs, and subsequently
produces enhanced acoustic tokens conditioned on both original and predicted
representations using another AR-LLM. Similarly, LauraTSE~\cite{tang2025lauratse}
adopts a two-stage framework based on LauraGPT~\cite{du2023lauragpt}.
LLaSE-G1~\cite{kang-etal-2025-llase} jointly models semantics and acoustics
within a single-stage pipeline, utilizing a Llama-style LLM backbone with a
single-layer codec (X-Codec2~\cite{ye2025codec,ye2025llasa}) under a
non-autoregressive strategy. These methods demonstrate the effectiveness of the
next-token prediction (NTP) paradigm in speech processing. Despite these advances,
existing LLM-based TSE methods remain restricted to audio-only scenarios.
To the best of our knowledge, TSE-Omni is the first LLM-based TSE system that
accepts visual cues as well as audio cues in one extractor backbone.

\subsection{Omni-LLMs}
Recently, multimodal LLMs (MLLMs) have demonstrated strong performance across
various speech understanding and generation tasks.
Qwen-Omni~\cite{xu2025qwen2,xu2025qwen3} employs a thinker-talker architecture
supporting flexible input modalities, while LongCat-Omni~\cite{team2025longcat}
and OmniVinci~\cite{ye2025omnivinci} utilize efficient cross-modal interaction
mechanisms for modalities such as speech, text, and video. These Omni-style MLLMs
leverage the foundational understanding and generative capabilities of LLMs to
unify diverse single-modality tasks, yielding significant performance gains.
Inspired by these advances, we propose TSE-Omni, which extends LLM-based
TSE to accept audio and
visual enrollment cues, targeting extraction in
complex real-world scenarios.

\section{Methods}

\subsection{TSE-Omni Overview}
\begin{figure*}[htbp]
 \centering
\includegraphics[scale=0.52]{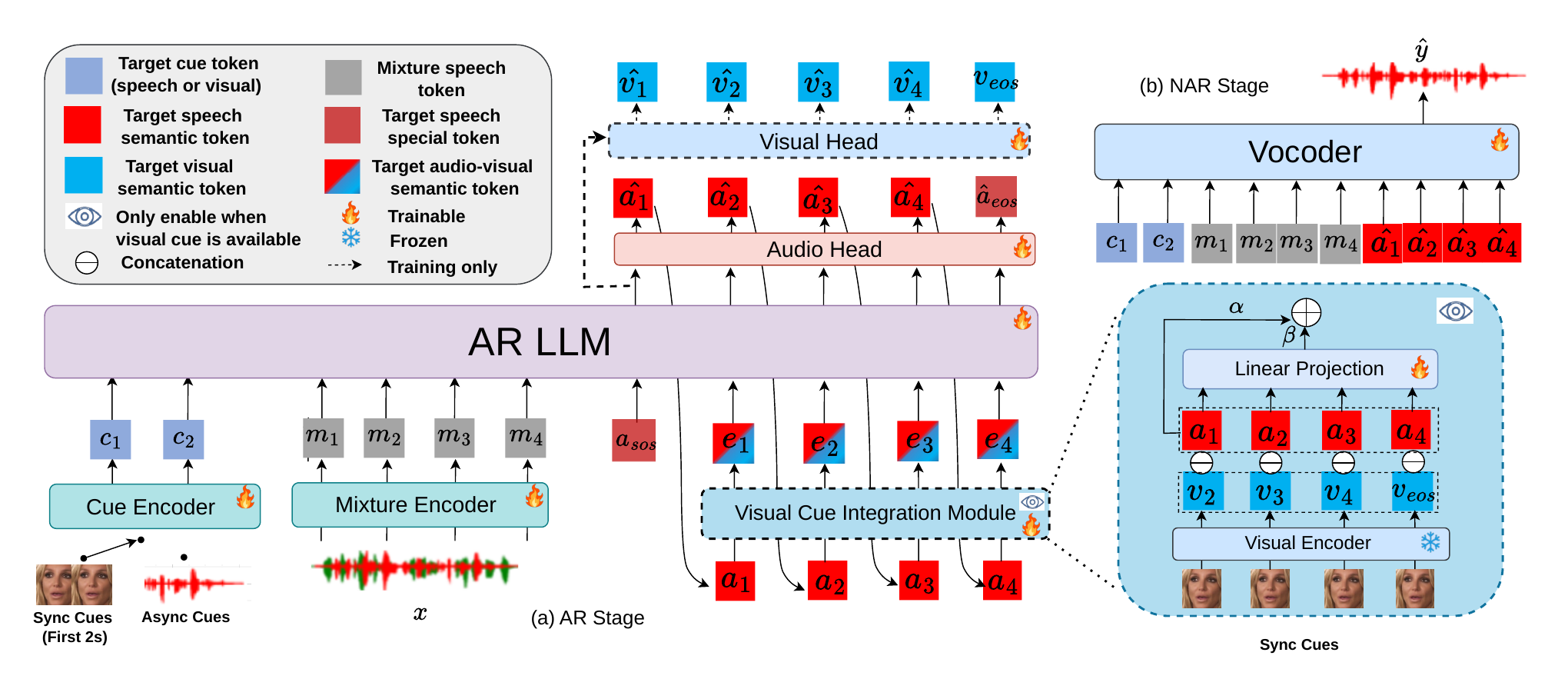}
 \caption{Overview of the TSE-Omni training framework. (a) \textbf{AR Stage}:
predicts target speech semantic tokens from the mixture, conditioned on
a pre-recorded audio clip or temporally synchronized visual cues. A short
initial visual segment (default 2\,s) goes to the cue encoder for recognition.
The visual integration module is used only for visual-cue TSE. After
$a_{\mathrm{sos}}$, visual tokens start at $v_2$: each $\hat{a}_t$ ($t\geq 2$)
is conditioned on $e_{t-1}=f_{av}(\hat{a}_{t-1},v_t)$. (b) \textbf{NAR Stage}:
reconstructs the waveform from the predicted semantic tokens, the mixture, and
the cue.}
 \label{fig:TSE-Omni-model}
\end{figure*}

Traditional TSE models handle audio and visual cues independently:
$\hat{y} = f_a(a, x)$ and $\hat{y} = f_v(v, x)$, where $f_a$ and $f_v$ denote
audio- and visual-cue models. This leads to redundant deployment and no
parameter sharing. TSE-Omni instead unifies both cues in a single AR-LLM
(Fig.~\ref{fig:TSE-Omni-model}), consisting of an autoregressive (AR) stage
that predicts target speech semantic tokens and a non-autoregressive (NAR)
stage that generates the waveform.

\textbf{AR Stage:}
The AR stage predicts discrete target speech semantic tokens $\hat{a}_{1:T}$,
taking the cue embedding $c$ and the mixture embedding $m$ as context, starting
from the special token $a_{\mathrm{sos}}$. The token-wise decoding condition
depends on the cue type and is detailed in Sec.~\ref{sec:self-enroll}, which
also introduces our self-enrollment and cross-modality compensation mechanism.

\textbf{NAR Stage:}
The NAR stage reconstructs the waveform $y$ from $\hat{a}_{1:T}$,
conditioned on $m$ and $c$. Following
\cite{du2023lauragpt,tang2025lauratse}, a six-layer Conformer predicts the
remaining acoustic tokens beyond the first $n_q$ RVQ layers, and a codec
decoder produces the speech signal. Both the NAR Conformer and the codec
decoder are trained.

\subsection{Mixture and Cue Encoders}
As shown in Fig.~\ref{fig:TSE-Omni-model}, the LLM's input at each AR step is a
continuous input sequence built from three continuous embeddings: the mixture
embedding $m$, the cue embedding $c$, and the embeddings of the previously
predicted speech tokens $\hat{a}_{1:t-1}$ (looked up from the first $n_q$
RVQ layers).

\textbf{Mixture Encoder.} A six-layer Conformer encoder processes the
mel-spectrogram of $x$ (the channel size is first projected to $D$) and produces the
mixture embedding $m$.

\textbf{Cue Encoder.} The figure shows a single \emph{cue encoder} block, which
abstracts a family of modality-specific encoders, one per cue type. Each
front-end produces a modality-specific feature, which is then projected to the
same $D$-dimensional space and encoded into a unified continuous cue embedding
$c$:
\begin{itemize}
    \item \textbf{Audio:} a pre-recorded target speech utterance is converted to a
    mel-spectrogram and encoded (sharing the mixture encoder weights).
    \item \textbf{Visual (Lip movement):} lip embeddings from the default VSR
    front-end are projected to $D$ (AV-HuBERT is compared in
    Sec.~\ref{sec:visual_tokens}).
    \item \textbf{Visual (Co-speech gesture):} a BLSTM
    encoder~\cite{co-speech-gestures} extracts gesture sequence embeddings
    (15\,Hz), which are then upsampled to the speech token frame rate (25\,Hz) by interpolation.
    \item \textbf{Text:} RoBERTa~\cite{Liu2019RoBERTaAR} extracts
    utterance-level text embeddings from partial target speech transcripts.
\end{itemize}
When multiple cues are given, their embeddings are concatenated and projected
to the shared space; missing modalities are replaced by zero tensors.

\textbf{Two Cue Roles.} A cue is either \emph{asynchronous} (pre-recorded
audio, text) or \emph{synchronous} (lip, gesture, temporally aligned with the
mixture) (Fig.~\ref{fig:cue_type}). This determines the role a cue plays
during extraction:

\begin{itemize}
    \item \textbf{Target-Speaker Recognition.} All cues perform recognition:
    they identify \emph{who} to extract at the cold start. Asynchronous cues
    are used directly; for synchronous cues, we use only a short initial
    segment (e.g., the first 2 seconds), which plays the same recognition
    role.
    \item \textbf{Target-Speaker Tracking.} Only synchronous cues can track,
    because they are temporally aligned with the mixture speech and can be fused with
    each newly predicted speech token $\hat{a}_{t-1}$ (from $t=2$ onward). For
    asynchronous cues, tracking is not possible via the cue; instead, it relies
    solely on the previously predicted speech tokens
    $\hat{a}_{1:t-1}$ via next-token prediction, as detailed in
    Sec.~\ref{sec:self-enroll}.
\end{itemize}

\subsection{Self-enrollment and Cross-modality Compensation}
\label{sec:self-enroll}

This is the core of our contribution: we exploit the LLM's NTP ability for audio-visual cue compensation. The AR stage
autoregressively predicts target speech semantic tokens
$\hat{a}_{1:T}$, where $t$ indexes the AR step
and $\hat{a}_t$ is the stack of the first $n_q$ RVQ code indices predicted at that step (the corresponding look-up embeddings are in $\mathbb{R}^{n_q \times D}$). Following
\cite{du2023lauragpt,tang2025lauratse}, we predict the first $n_q = 2$ of the
$32$ RVQ layers.

For \emph{audio-cue} TSE, $\hat{a}_t$ is conditioned only on the historical
speech tokens $e_{1:t-1} = \hat{a}_{1:t-1}$; the visual module is inactive.
For \emph{visual-cue} TSE, the prediction is conditioned on historical
audio-visual joint tokens
\begin{equation}
\begin{aligned}
e_{1:t-1} &= f_{av}\bigl(\hat{a}_{1:t-1},\, v_{2:t}\bigr) \\
          &= \alpha\, \hat{a}_{1:t-1}
             + \beta\, \mathrm{Linear}\bigl(\mathrm{concat}
             (\hat{a}_{1:t-1},\, v_{2:t})\bigr),
\end{aligned}
\label{eq:fav}
\end{equation}
where the residual module $f_{av}$ in Eq.~\eqref{eq:fav} operates
independently on each token pair. Concretely, each integrated token $e_{t-1}$
concatenates the look-up embedding of the last predicted token $\hat{a}_{t-1}$
with the synchronized visual token $v_t$ along the channel dimension, and
passes the result through $f_{av}$. Since $\hat{a}_1$ has no preceding token to
fuse with, the visual tokens start at $v_2$; each $\hat{a}_t$ is thus conditioned on $e_{t-1}$, a fusion of the previous
speech token $\hat{a}_{t-1}$ and the visual frame $v_t$.

The residual design of $f_{av}$ keeps the speech token (weight $\alpha$) while
fusing in the projected audio-visual token (weight $\beta$). This mitigates
modality imbalance, since the LLM backbone is primarily adapted to speech
tokens. At inference, later corrupted or missing visual frames are replaced by
zeros; the visual module stays on. The AR stage still conditions on
$\hat{a}_{1:t-1}$ (self-enrollment), so the $\beta$ branch sees
$\mathrm{concat}(\hat{a},\,0)$ rather than a learned ignore-visual gate.
This is the protocol used in the visual-corruption experiments: no
corruption-matched training and no extra recovery module. The residual
$\alpha\,\hat{a}$ path is what keeps extraction going after the 2\,s clean
visual start.

To signal the end of the sequence, we predict the end-of-sequence token
$a_{\mathrm{eos}}$ using a zero embedding for the corresponding visual token,
avoiding an extra special token.

\subsection{Training}
During core Omni-training, half of each batch is assigned to audio-cue TSE and
half to visual-cue TSE (lip movements on VoxCeleb2). Text and co-speech gesture
cues are trained later on their own mixtures (Sec.~\ref{sec:datasets}). The
default loss, used for all main results, is
\[
\mathcal{L} = \mathcal{L}_{CE}(\hat{a}, a) + \mathcal{L}_{MSE}(\hat{a}_{emb}, a_{emb}),
\]
where $\hat{a}$ and $a$ are predicted and ground-truth discrete speech semantic
tokens, and $\hat{a}_{emb}$ and $a_{emb}$ are predicted and ground-truth full
target speech embeddings (this MSE trains the NAR Conformer and codec decoder).
An optional visual-head loss $\mathcal{L}_{CE}(\hat{v}, v)$ on discrete visual
tokens is studied only in the visual-token ablation of
Sec.~\ref{sec:visual_tokens}; it is not used in the default model.

\section{Implementation}

The target speaker is identified by the cue-encoder branch, which
recognizes the target identity, while the tracking branch provides
frame-level semantic features. Both the speech and visual tokens share a
dimension of $D = 128$ and are sampled at 25 Hz.

Training proceeds in two stages. Before the Omni-training stage, the audio
cue encoder and the visual cue encoder (excluding its tracking branch) are
first aligned with an MSE loss; skipping this alignment leads to clearly
degraded performance when the visual cue is used. During the Omni-training
stage, the learning rate starts at $10^{-3}$ and is halved whenever the
validation loss does not improve for 3 consecutive epochs, and training
stops after 6 consecutive epochs without improvement.

\textbf{Lip Movement Encoder and Visual Head.} We use a VSR (ResNet)~\cite{vsr_tpami} front-end to
extract visual semantic tokens; it consists of a 3D convolutional layer
(Conv3D) followed by a ResNet18 block, pretrained on LRS3 with the lip-reading
task. The default model does not attach a visual prediction head. For the
optional auxiliary loss in Sec.~\ref{sec:visual_tokens} only, we
follow the data-preprocessing protocol of AV-HuBERT
Large~\cite{AVhubert} to obtain discrete visual tokens, feeding only visual data
into AV-HuBERT.

\textbf{Metrics.} We use SpeechBERTScore (SBS)~\cite{saeki2024speechbertscore}
to evaluate the semantic similarity between ground-truth and extracted
speech. Speaker similarity (SIM) is measured by the WavLM-based similarity
metric~\cite{WavLM}, while speech quality is evaluated with NISQA~\cite{NISQ}
and DNSMOS~\cite{DNSMOS}. All metrics are widely adopted in generative SE
studies~\cite{gense,kang-etal-2025-llase}; higher is better. Streaming latency is measured by the real-time factor (RTF).

\section{Experimental Setups}

\subsection{Datasets}
\label{sec:datasets}

\textbf{Core Dataset.}
We use VoxCeleb2~\cite{voxceleb2} as our primary training and in-domain
evaluation set: its two-speaker mixtures are more challenging, noisier, and
much closer to real-world conditions than LRS3, and it provides audio, visual
(lip-movement), and transcript cues for the same utterances. We simulate a two-speaker core
dataset with $400{,}000$/$2{,}000$/$2{,}000$ utterances for
training/validation/test, respectively (around $400$ h of training data). The
scale roughly matches the $460$ h of LibriSpeech used to pretrain the LauraTSE
backbone. That hour count is a data-budget reference only: on Libri2mix we
evaluate TSE-Omni zero-shot, whereas LauraTSE is trained in-domain.
Unless a table names another training set, TSE-Omni numbers
use this VoxCeleb2 core checkpoint. For audio-cue-based TSE, we randomly select an utterance
from the target speaker as a pre-recorded audio cue. All utterances are randomly
clipped between $3$ and $6$ seconds, and the default cue duration is 2
seconds for both audio and visual cues. The interfering speaker is mixed at a
signal-to-interference ratio (SIR) randomly sampled between $-5$~dB and $5$~dB.

\textbf{LRS3 Evaluation.}
LRS3 mixtures provide both visual and audio cues. Visual-cue baselines on LRS3
(USEV, AV-Sepformer, AV-Mamba, AVDiffuSS, FlowAVSE) are trained in-domain.
Table~\ref{tab:unified_compact_full} and the LRS3 ablations test the VoxCeleb2
core checkpoint on two-speaker LRS3 mixtures simulated with the same duration
and SIR protocol: TSE-Omni (V) and TSE-Omni (A) are both zero-shot. We do not
train TSE-Omni on LRS3, except for the target-switch fine-tuning set below.
The VSR front-end remains the LRS3-pretrained lip-reading encoder described in
Implementation.

\textbf{Libri2mix Evaluation.}
Libri2mix rows in Table~\ref{tab:unified_compact_full} compare in-domain audio
baselines with a zero-shot TSE-Omni. NeMo, LLaSE-G1, LauraTSE, and SoloSpeech
are trained on Libri2mix (or the LibriSpeech/Libri2mix setup of the original
papers). TSE-Omni (A) is the VoxCeleb2 core checkpoint, not trained on
Libri2mix. TSE-Omni (A+Pretrain) is initialized from the official LauraTSE
checkpoint and then fine-tuned on VoxCeleb2; it is not a Libri2mix-trained
TSE-Omni, but it is not zero-shot either, because LauraTSE saw Libri2mix.

\begin{figure}[htbp]
\centering
\includegraphics[scale=0.8]{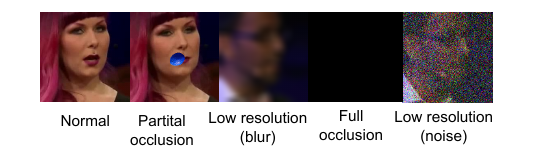}
\caption{Visual frame corruption on the VoxCeleb2 evaluation protocol:
full missing, partial occlusion, and low resolution. The first 2\,s stay
clean; later corrupted or missing frames are zeros.}
\label{fig:vocc_EXAMPLE}
\end{figure}

\textbf{Visual-Cue-Impaired Test Set.}
We apply three corruption types to the VoxCeleb2 core test set
($2{,}000$ utterances): full missing, partial occlusion, and low resolution
(Fig.~\ref{fig:vocc_EXAMPLE}). To ensure a stable start, the first 2 seconds of
each visual cue are kept clean, and corrupted or missing later frames are
replaced by zeros; the visual module stays on. The full-missing protocol zeros
every visual frame after 2\,s (the target leaving the view). Full-missing and
the three-type average are reported in Sec.~\ref{sec:speech_compensate} on this
same $2{,}000$-utterance core test set.

\textbf{Target-Speaker-Switch Dataset.}
We randomly select two target utterances of at least 4 seconds (before and
after the switch) from LRS3 and simulate $20{,}000$/$1{,}000$ mixtures for
fine-tuning and test, respectively. In each sample,
the target speaker switches mid-utterance at a randomly selected timestamp
between 3 and 3.5 seconds. To keep acoustic interference consistent, we
randomly select an interfering utterance of at least 6 seconds. This model
is fine-tuned from the VoxCeleb2 core checkpoint; it is the only LRS3-trained
extractor in this paper. Switching is evaluated with visual cues (clean or
occluded), not with audio-only dual enrollment.

\textbf{Three-Speaker Test Set.}
We construct a 3-mix test set from VoxCeleb2~\cite{voxceleb2} to study
the model's robustness under multi-talker interference. Each mixture
contains one target speaker and two interfering speakers, with SIR sampled from $[-10,10]$~dB. The test set consists of $3{,}000$ utterances.
The simulated mixtures have variable durations, randomly between $4$--$6$ seconds.
This test set is used only for
zero-shot evaluation with the model trained on the core training set.

\textbf{Sparse-Overlap Dataset.}
We construct a sparse-overlap dataset from IEMOCAP~\cite{iemocap} to study
extraction in conversation-style mixture speech: target and interfering
utterances are mixed with large silences between the two talkers, while visual
inputs are kept clean. The dataset consists of $400{,}000$/$9{,}621$/$5{,}528$ utterances
for training/validation/test. This isolates the model's ability to predict
silent speech tokens from visual cues (e.g., lip non-movement) or pre-enrolled
audio cues. The overlap ratio follows USEV~\cite{usev}, with SIR sampled
from $[-5,5]$~dB. The simulated mixtures have variable durations up to 10
seconds, and we use 6-second inputs for inference. This model is fine-tuned from
the checkpoint pretrained on the core training set.

\textbf{Cue-Combination Dataset.}
To study modality extension and combination, we derive additional cues on
VoxCeleb2~\cite{voxceleb2} and the YGD dataset~\cite{co-speech-gestures}. VoxCeleb2
provides lip movements, transcripts, and pre-recorded audio; YGD provides
co-speech gesture sequences, transcripts, and pre-recorded audio. For both
datasets, partial transcriptions are obtained by randomly masking out $20\%$
of the full transcription, extracted with Whisper~V2~\cite{whisper}.
For YGD with co-speech gesture cues, we follow the SEG network~\cite{co-speech-gestures}:
we build YGD-2mix from the YouTube gesture
dataset~\cite{co-speech-gestures} of $1{,}696$ TED videos
($27{,}611$/$3{,}654$/$3{,}475$ train/validation/test segments). Following the
SEG pipeline, 2D poses are lifted to 3D poses at 15 fps as co-speech gestures;
we simulate $200{,}000$/$5{,}000$/$3{,}000$ mixtures for the training, validation,
and test sets at a random interferer SIR in $[-10,10]$~dB, with 16 kHz audio.
VoxCeleb2 cue-combination models are trained from scratch on VoxCeleb2;
YGD models are trained from scratch on YGD-2mix.  

\subsection{Baselines}
We compare TSE-Omni against state-of-the-art baselines, encompassing both
generative and discriminative paradigms.

\textbf{Audio-Cue Baselines:}
We benchmark against SoloSpeech~\cite{solospeech}, a DiT-based model with a
flow-matching objective; NeMo~\cite{NEMO_SSL}, a generative model utilizing
STFT-based spectral processing; LLaSE-G1~\cite{kang-etal-2025-llase}, a
Llama-based TSE model employing discrete tokens and non-autoregressive
decoding; and LauraTSE~\cite{tang2025lauratse}, a decoder-only LLM streamlined
from LauraGPT~\cite{du2023lauragpt} for efficient speech extraction.

\textbf{Visual-Cue Baselines:}
We include USEV~\cite{usev} and AV-Sepformer~\cite{av-sepformer}, representing
RNN-based and Transformer-based architectures respectively, alongside
AV-Mamba~\cite{wu2026elegance}, which leverages state-space models for
efficiency. For generative approaches, we compare with
AVDiffuSS~\cite{avdiffus}, a two-stage pipeline combining discriminative
extraction with generative refinement, and with its flow-matching variant
FlowAVSE~\cite{FLOWAVSE}.

\begin{table*}[t!]
\centering
\scriptsize
\TSETabSetup
\begin{threeparttable}
\caption{Comprehensive performance benchmarking across LRS3, VoxCeleb2, and
Libri2mix. LRS3 supports both visual and audio cues. Visual-cue baselines on
LRS3 are trained in-domain; TSE-Omni (V) and TSE-Omni (A) LRS3 rows are
zero-shot from the VoxCeleb2 core checkpoint. VoxCeleb2 rows are in-domain.
On Libri2mix, audio baselines are
trained in-domain; TSE-Omni (A) is zero-shot from the VoxCeleb2 core
checkpoint; TSE-Omni (A+Pretrain) is initialized from LauraTSE and fine-tuned
on VoxCeleb2. On LRS3 and VoxCeleb2, the best \emph{visual-cue} value per
metric is in bold (audio-cue TSE-Omni is listed for reference). On Libri2mix,
the best value per metric is in bold. Higher is better. \textbf{Param.:} speech extractor backbone parameters;
\textbf{Arch.:} speech extractor backbone architecture; \textbf{Obj.:} prediction
objective; \textbf{Inp.~rep.:} intermediate modeling representation; \textbf{Rate:} frame
rate; \textbf{Type:} D = discriminative, G = generative; \textbf{Cue:} A =
audio, V = visual.}
\label{tab:unified_compact_full}
\begin{tabular}{@{} l l c c c c c c c *{6}{Y} @{}}
\toprule
\multirow{2}{*}{\textbf{Test set}} & \multirow{2}{*}{\textbf{Model}}
 & \multirow{2}{*}{\textbf{Param}} & \multirow{2}{*}{\textbf{Arch.}}
 & \multirow{2}{*}{\textbf{Obj.}} & \multirow{2}{*}{\textbf{Inp. rep.}}
 & \multirow{2}{*}{\textbf{Rate}} & \multirow{2}{*}{\textbf{Type}}
 & \multirow{2}{*}{\textbf{Cue}}
 & \multicolumn{3}{c}{\textbf{Semantic-Speaker-Quality$\uparrow$}}
 & \multicolumn{3}{c}{\textbf{DNSMOS$\uparrow$}} \\
\cmidrule(lr){10-12} \cmidrule(lr){13-15}
 & & & & & & & & &\textbf{SBS} & \textbf{SIM} & \textbf{NISQA} & \textbf{SIG} & \textbf{BAK} & \textbf{OVL} \\
\midrule
\multirow{7}{*}{\textbf{LRS3}}
& USEV~\cite{usev} & 16.4M & RNN & Mask & Emb & 800 Hz & D & V & 0.83 & \textbf{0.96} & 2.95 & 3.24 & 3.61 & 2.80 \\
& AV-Sepf~\cite{av-sepformer} & 163M & Sepf. & Mask & Emb & 2000 Hz & D & V & 0.87 & \textbf{0.96} & 2.46 & 3.19 & 2.90 & 2.43 \\
& AV-Mamba~\cite{wu2026elegance} & 126M & Mamba & Mask & Emb & 800 Hz & D & V & \textbf{0.89} & 0.95 & 2.82 & 3.23 & 3.34 & 2.66 \\
& AVDiffuSS~\cite{avdiffus} & 241M & U-Net & Noise & STFT & 125 Hz & D+G & V & 0.65 & 0.87 & 3.18 & 3.35 & 3.45 & 2.78 \\
& FlowAVSE~\cite{FLOWAVSE} & 241M & U-Net & Vel. & STFT & 125 Hz & D+G & V & 0.74 & 0.93 & 2.66 & 3.31 & 2.89 & 2.51 \\
& \textbf{TSE-Omni (V)} & 77M & AR-LLM & RVQ & RVQ & 25 Hz & G & V & \textbf{0.89} & \textbf{0.96} & \textbf{3.84} & \textbf{3.51} & \textbf{3.76} & \textbf{3.08} \\
& \textbf{TSE-Omni (A)} & 77M & AR-LLM & RVQ & RVQ & 25 Hz & G & A &
0.91 & 0.97 & 3.89 & 3.50 & 3.79 & 3.08 \\
\midrule
\multirow{7}{*}{\textbf{VoxCeleb2}}
& USEV~\cite{usev} & 16.4M & RNN & Mask & Emb & 800 Hz & D & V & 0.80 & 0.94 & 2.55 & 3.18 & 3.29 & 2.60 \\
& AV-Sepf~\cite{av-sepformer} & 163M & Sepf. & Mask & Emb & 2000 Hz & D & V & \textbf{0.81} & \textbf{0.95} & 2.08 & 3.15 & 2.65 & 2.29 \\
& AV-Mamba~\cite{wu2026elegance} & 126M & Mamba & Mask & Emb & 800 Hz & D & V & 0.79 & 0.90 & 2.50 & 3.17 & 2.91 & 2.43 \\
& AVDiffuSS~\cite{avdiffus} & 241M & U-Net & Noise & STFT & 125 Hz & D+G & V & 0.60 & 0.74 & 2.89 & 3.21 & 2.89 & 2.46 \\
& FlowAVSE~\cite{FLOWAVSE} & 241M & U-Net & Vel. & STFT & 125 Hz & D+G & V & 0.62 & 0.76 & 2.69 & 3.11 & 2.96 & 2.46 \\
& \textbf{TSE-Omni (V)} & 77M & AR-LLM & RVQ & RVQ & 25 Hz & G & V & \textbf{0.81} & 0.94 & \textbf{3.24} & \textbf{3.48} & \textbf{3.35} & \textbf{2.85} \\
& \textbf{TSE-Omni (A)} & 77M & AR-LLM & RVQ & RVQ & 25 Hz & G & A & 0.79 & 0.93 & 3.23 & 3.47 & 3.33 & 2.84 \\
\midrule
\multirow{6}{*}{\textbf{Libri2mix}}
& NeMo~\cite{NEMO_SSL} & 1.7B & U-DIT & Vel. & STFT & 125 Hz & G & A & 0.53 & 0.85 & 2.21 & 3.29 & 3.13 & 2.61 \\
& LLaSE-G1$^{*}$~\cite{kang-etal-2025-llase} & 3.2B & AR-LLM$^{*}$ & VQ & Emb & 50 Hz & G & A & 0.73 & 0.90 & 3.89 & 3.30 & 3.94 & 2.99 \\
& LauraTSE~\cite{tang2025lauratse} & 77M & AR-LLM & RVQ & RVQ & 25 Hz & G & A & \textbf{0.90} & \textbf{0.97} & \textbf{4.01} & \textbf{3.60} & \textbf{4.06} & 3.21 \\
& SoloSpeech fast$^{\dagger}$ & 513M & U-DIT & Vel. & STFT-VAE & 50 Hz & G & A & 0.88 & \textbf{0.97} & 3.94 & 3.55 & 4.02 & 3.25 \\
& \textbf{TSE-Omni (A)} & 77M & AR-LLM & RVQ & RVQ & 25 Hz & G & A & 0.83 & 0.94 & 3.69 & 3.57 & \textbf{4.06} & \textbf{3.29} \\
& \textbf{TSE-Omni (A+Pretrain)}$^{\ddagger}$ & 77M & AR-LLM & RVQ & RVQ & 25 Hz & G & A & 0.88 & \textbf{0.97} & 3.98 & 3.58 & 4.05 & 3.28 \\
\bottomrule
\end{tabular}
\begin{tablenotes}[flushleft]
\footnotesize
\item $^{*}$LLaSE-G1 uses non-autoregressive decoding.\quad
$^{\dagger}$SoloSpeech fast is the version without the refine stage.\quad
$^{\ddagger}$TSE-Omni (A+Pretrain) is initialized from LauraTSE (Libri2mix) and fine-tuned on VoxCeleb2, then tested on Libri2mix. TSE-Omni (A) is zero-shot from the VoxCeleb2 core checkpoint. Libri2mix baselines are trained in-domain.
\end{tablenotes}
\end{threeparttable}
\end{table*}

\section{Results and Analysis}

\subsection{Comparison with Other Backbones}
In this section, we evaluate TSE-Omni against widely used
audio-cue-based and visual-cue-based TSE baselines on the VoxCeleb2 core test
set and, zero-shot, on LRS3 (visual and audio cues) and on Libri2mix (audio only).
As shown in Table~\ref{tab:unified_compact_full}, on the LRS3 test set
(in-domain visual baselines; TSE-Omni zero-shot from the VoxCeleb2 checkpoint),
TSE-Omni (V) achieves an SBS of 0.89, tying the strongest visual-cue baseline
(AV-Mamba), while delivering the highest speech quality among all compared
models (NISQA 3.84). On VoxCeleb2 (Vox2), our primary evaluation set, TSE-Omni (V) matches
the best visual baseline in semantic similarity (SBS 0.81, tied with
AV-Sepformer) while attaining markedly higher speech quality (NISQA 3.24, vs.
2.55/2.08/2.50 for USEV/AV-Sepformer/AV-Mamba, respectively). Similar trends
hold for speaker similarity (SIM) and the DNSMOS metrics (SIG, BAK, OVL),
indicating that TSE-Omni successfully incorporates visual cues into an
audio-oriented LLM-TSE architecture while largely retaining its audio-cue
capability (Vox2 audio-cue SBS 0.79).

For audio cues on Libri2mix, the baselines are trained in-domain, while
TSE-Omni (A) is zero-shot from the VoxCeleb2 checkpoint. TSE-Omni (A) reaches
an SBS of 0.83, below LauraTSE (0.90), but still records the best overall
quality (OVL 3.29, above LauraTSE's 3.21). Initializing from LauraTSE and
fine-tuning on VoxCeleb2 (TSE-Omni (A+Pretrain)) is no longer zero-shot on
this set; it raises SBS from 0.83 to 0.88, closing most of the semantic gap to
in-domain LauraTSE while keeping OVL 3.28.

A notable distinction lies in the compression rate of the input speech
representations. TSE-Omni employs a 25 Hz frame rate for both speech and visual
semantic tokens, whereas most baselines utilize a more fine-grained
representation with lower compression rates. This suggests that during the
extraction stage, TSE-Omni primarily operates at the semantic level, which
makes it reasonable to align speech and visual semantic tokens at this stage,
while other baselines place greater emphasis on acoustic details.

\subsection{Speech Token Compensation}
\label{sec:speech_compensate}

\begin{figure}[htbp]
\centering
\includegraphics[width=1\linewidth]{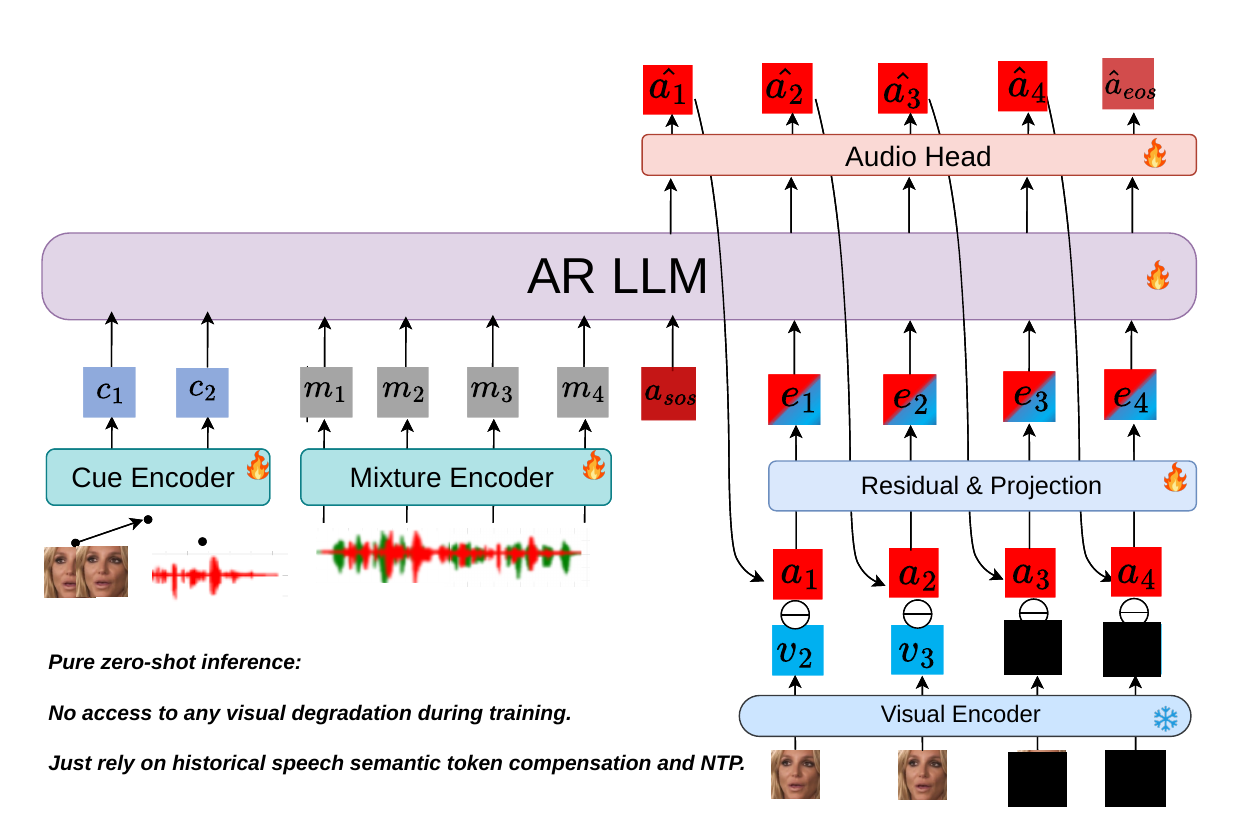}
\caption{Speech semantic token compensation. Only the first 2\,s of visual
cues are available; later visual frames are zeros. TSE-Omni continues
extraction from the historical predicted speech semantic tokens
(self-enrollment), with the visual module still enabled.}
\label{fig:speech_compensate}
\end{figure}

The self-enrollment design gives TSE-Omni a unique advantage: since every AR
prediction is conditioned on the previously predicted target speech semantic
tokens, the model can fall back on this accumulated speech-semantic context
to keep tracking the target speaker when later visual frames are zero-filled
(Fig.~\ref{fig:speech_compensate}). This protocol does not use
corruption-matched training data or a recovery module; corrupted frames are
zeros, and the visual module stays on.

We first validate compensation under the most challenging setting: full
missing, where only the visual cues of the first 2\,s are available and later
frames are zeros, mimicking a target speaker who moves out of view.
As shown in Table~\ref{tab:model_comparison_clean}, baselines that rely on
continuously available visual cues degrade severely: USEV and AV-Sepformer
drop to SBS of 0.71 and 0.70, respectively, and AV-Sepformer's NISQA
collapses to 1.96. In contrast, TSE-Omni (V) maintains an SBS of 0.81, a
SIM of 0.95, and a NISQA of 3.06 on this same $2{,}000$-utterance test set
as the clean VoxCeleb2 evaluation, confirming that a 2\,s visual cold start
plus speech-token self-enrollment is sufficient to sustain extraction.

\begin{table}[!t]
\centering
\footnotesize
\TSETabSetup
\caption{VoxCeleb2 core test set ($2{,}000$ utterances), full-missing visual
cues: first 2\,s clean, later visual frames zeros. Same mixtures as the clean
VoxCeleb2 test in Table~\ref{tab:unified_compact_full}.}
\label{tab:model_comparison_clean}
\begin{tabular}{@{} lc *{6}{Y} @{}}
\toprule
\multirow{2}{*}{\textbf{Model}} & \multirow{2}{*}{\textbf{Cue}}
 & \multicolumn{3}{c}{\textbf{Semantic-Speaker-Quality $\uparrow$}}
 & \multicolumn{3}{c}{\textbf{DNSMOS $\uparrow$}} \\
\cmidrule(lr){3-5} \cmidrule(lr){6-8}
 & & \textbf{SBS} & \textbf{SIM} & \textbf{NISQA} & \textbf{SIG} & \textbf{BAK} & \textbf{OVL} \\
\midrule
USEV~\cite{usev}             & V & 0.71 & 0.86 & 2.56 & 3.17 & 2.94 & 2.43 \\
AV-Sepf~\cite{av-sepformer} & V & 0.70 & 0.84 & 1.96 & 3.04 & 2.54 & 2.21 \\
AV-Mamba~\cite{wu2026elegance} & V & 0.79 & 0.86 & 2.42 & 3.14 & 2.89 & 2.39 \\
AVDiffuSS~\cite{avdiffus}    & V & 0.60 & 0.74 & 2.89 & 3.21 & 2.89 & 2.46 \\
FlowAVSE~\cite{FLOWAVSE}     & V & 0.62 & 0.76 & 2.69 & 3.11 & 2.96 & 2.46 \\
\midrule
\textbf{TSE-Omni (V)}        & V & \textbf{0.81} & \textbf{0.95} & \textbf{3.06} & \textbf{3.46} & \textbf{3.32} & \textbf{2.84} \\
\bottomrule
\end{tabular}
\end{table}

\begin{figure}[htbp]
\centering
\includegraphics[width=\linewidth]{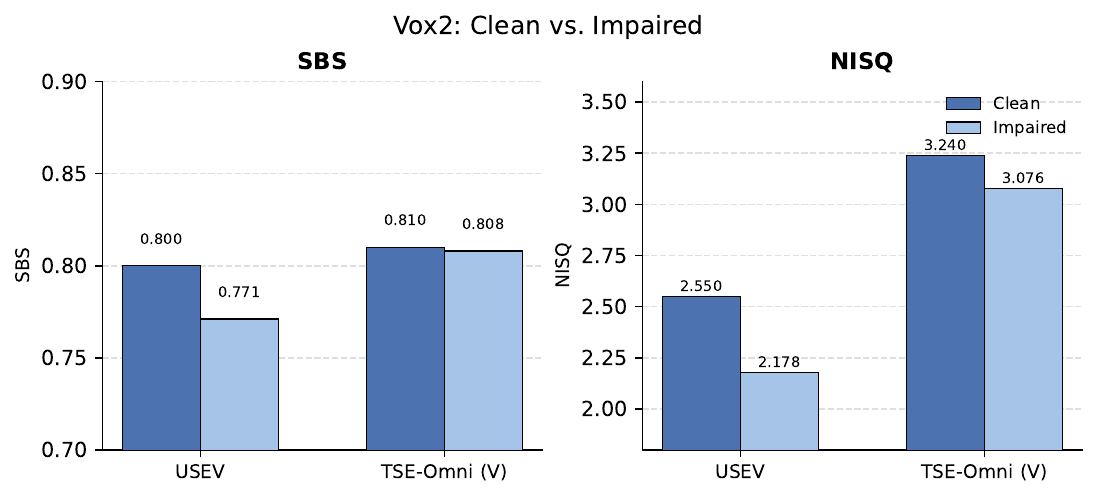}
\caption{VoxCeleb2, TSE-Omni~(V) vs.\ USEV: clean visual cues versus the
average of full missing, partial occlusion, and low resolution (2\,s clean
start, later frames zero-filled). Metrics are SBS and NISQA.}
\label{fig:clean_vs_impaired}
\end{figure}

We then compare TSE-Omni~(V) with USEV under clean and impaired visual conditions
on the VoxCeleb2 test set, with results in Fig.~\ref{fig:clean_vs_impaired}.
Two conclusions emerge.
\textbf{First}, TSE-Omni~(V) consistently outperforms USEV under both
conditions and on both metrics: on SBS it reaches $0.810$ (clean) and
$0.808$ (impaired), against USEV's $0.800$ and $0.781$; on NISQA the gap is
even larger, $3.24$ and $3.08$ versus $2.55$ and $2.18$. Notably, even
under impaired conditions, TSE-Omni's NISQA ($3.08$) stays well above USEV's
\emph{clean} NISQA ($2.55$).
\textbf{Second}, self-enrollment makes TSE-Omni far less sensitive to
visual degradation: from clean to impaired, USEV's NISQA drops by $0.37$
and its SBS by $0.019$, whereas TSE-Omni degrades by only $0.16$ and
$0.002$, respectively; SBS on the core test set is effectively unchanged when
later visual frames are zero-filled, while NISQA still drops (3.24 to 3.08).

More importantly, real-world visual degradation is open-ended: occlusion,
low resolution, motion blur, poor lighting (Fig.~\ref{fig:vocc_EXAMPLE}), or the speaker leaving the
frame cannot be exhaustively anticipated. This work considers three
representative types, namely full missing, partial occlusion, and low
resolution, and the impaired results in
Fig.~\ref{fig:clean_vs_impaired} are their average.

Existing visual-cue
TSE models often need corruption-matched training or a recovery module, or they
degrade sharply (Table~\ref{tab:model_comparison_clean}). TSE-Omni instead zeros
later visual frames and relies on self-enrollment; we do not train on simulated
corruptions or add a visual-recovery module.

\subsection{Challenging Scenarios}

\subsubsection{Target Speaker Switch}

\label{sec:switch}
\begin{figure}[htbp]
\centering
\includegraphics[width=1.1\linewidth]{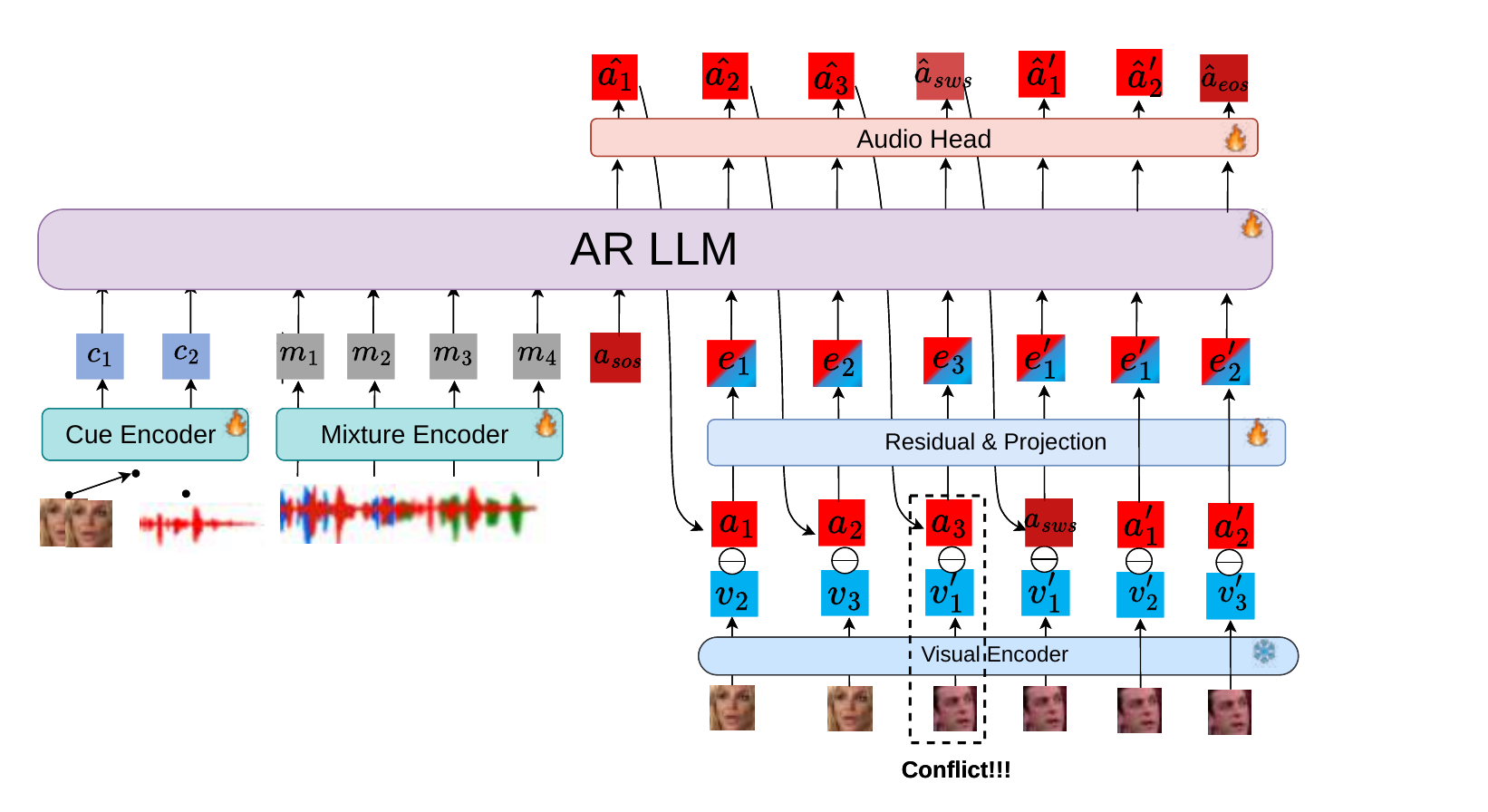}
\caption{Target-speaker switch on a visual cue: a switch token (SWS) is added
to the vocabulary. The LLM detects a conflict between the new visual cue and
the self-enrolled speech history, resets the AR context, and follows the new
target.}
\label{fig:target_switch}
\end{figure}

We evaluate TSE-Omni on the target speaker switching scenario, where the
target speaker changes mid-utterance, under two visual conditions: clean
(vclean) and occluded (vocc), as illustrated in
Fig.~\ref{fig:target_switch}. Results are reported in
Table~\ref{tab:switch}.

\begin{table}[h]
\centering
\footnotesize
\TSETabSetup
\caption{Target speaker switch on LRS3 (fine-tuned from the VoxCeleb2 core
checkpoint). ``SWS'' is the switch token. We report DNSMOS-OVL only; SIG/BAK
follow the same trend. Best values within each visual condition are bold.}
\label{tab:switch}
\begin{tabular}{@{} ll c *{4}{Z} @{}}
\toprule
\multirow{2}{*}{\textbf{Method}} & \multirow{2}{*}{\textbf{Visual}}
 & \multirow{2}{*}{\textbf{SWS}}
 & \multicolumn{3}{c}{\textbf{Sem.-Spk.-Qual$\uparrow$}}
 & \textbf{DNSMOS$\uparrow$} \\
\cmidrule(lr){4-6} \cmidrule(lr){7-7}
 & & & \textbf{SBS} & \textbf{SIM} & \textbf{NISQA} & \textbf{OVL} \\
\midrule
\multirow{1}{*}{USEV} & \multirow{3}{*}{vclean} & -- & 0.808 & \textbf{0.949} & 2.635 & 2.744 \\
\cmidrule(lr){3-7}
\multirow{2}{*}{TSE-Omni (V)} & & w/o & 0.774 & 0.928 & 3.495 & 3.082 \\
 & & w/  & \textbf{0.812} & 0.936 & \textbf{3.560} & \textbf{3.120} \\
\midrule
\multirow{1}{*}{USEV} & \multirow{3}{*}{vocc} & -- & 0.736 & 0.926 & 2.477 & 2.650 \\
\cmidrule(lr){3-7}
\multirow{2}{*}{TSE-Omni (V)} & & w/o & 0.766 & 0.925 & 3.474 & 3.074 \\
 & & w/  & \textbf{0.809} & \textbf{0.934} & \textbf{3.560} & \textbf{3.140} \\
\bottomrule
\end{tabular}
\end{table}
Two observations emerge from Table~\ref{tab:switch}. \textbf{First},
TSE-Omni with the switch token (w/ SWS) consistently outperforms USEV on
all quality metrics under both conditions, and the advantage is larger
under vocc, where visual cues are degraded (SBS 0.809 vs.\ 0.736). USEV
retains the highest SIM under vclean (0.949), since masking-based
extraction preserves the timbre of the tracked speaker, but it suffers
severe quality degradation (NISQA 2.635). \textbf{Second}, incorporating
SWS yields consistent gains over its counterpart without the token (w/o
SWS): SBS improves from 0.774 to 0.812 under vclean and from 0.766 to
0.809 under vocc (an absolute gain of about $0.04$), confirming that the
switch token is the key to handling speaker transitions.

These results indicate that the LLM backbone successfully learns to
predict the SWS special token at the exact switching point, triggered by a
conflict between the new target speaker's visual cue and the historical
speech-token prediction. This mechanism effectively resets the accumulated
self-enrollment history and mitigates inter-speaker interference during
transitions.

\subsubsection{Three-Speaker}

Table~\ref{tab:3spk} presents the zero-shot performance comparison on the
VoxCeleb2 test set under three-speaker interference
(one target, two interferers).
TSE-Omni (A) is best on every reported metric. TSE-Omni (V) matches USEV on
SBS (0.629 vs.\ 0.626) and is much higher on quality (NISQA 2.820 vs.\ 1.774),
but its SIM (0.867) is below USEV (0.879). These findings highlight
generalization under extra interferers, with audio enrollment remaining the
stronger identity cue in this 3-mix setting.

\begin{table}[h]
\centering
\footnotesize
\TSETabSetup
\caption{Three-speaker mixtures on the VoxCeleb2 test set (zero-shot from the
core checkpoint). Best values within each metric are bold.}
\label{tab:3spk}
\begin{tabular}{@{} l *{6}{Z} @{}}
\toprule
\multirow{2}{*}{\textbf{Model}}
 & \multicolumn{3}{c}{\textbf{Semantic-Speaker-Quality$\uparrow$}}
 & \multicolumn{3}{c}{\textbf{DNSMOS$\uparrow$}} \\
\cmidrule(lr){2-4} \cmidrule(lr){5-7}
 & \textbf{SBS} & \textbf{SIM} & \textbf{NISQA} & \textbf{SIG} & \textbf{BAK} & \textbf{OVL} \\
\midrule
USEV~\cite{usev}      & 0.626 & 0.879 & 1.774 & 3.037 & 2.327 & 2.087 \\
\textbf{TSE-Omni (V)} & 0.629 & 0.867 & 2.820 & 3.393 & 2.819 & 2.513 \\
\textbf{TSE-Omni (A)} & \textbf{0.658} & \textbf{0.898} & \textbf{2.827} & \textbf{3.395} & \textbf{2.843} & \textbf{2.535} \\
\bottomrule
\end{tabular}
\end{table}

\subsubsection{Sparse Overlap}

We evaluate extraction under conversation-style sparse overlap on
IEMOCAP~\cite{iemocap}, where the two talkers alternate with long silences
and visual cues are kept clean. As shown in Table~\ref{tab:sparse},
TSE-Omni (V) and TSE-Omni (A) perform comparably: they are slightly lower
than USEV on SBS and SIM, but clearly better on all quality metrics (NISQA
1.52 vs.\ 1.30; OVL 2.26 vs.\ 1.98).

\begin{table}[h]
\centering
\footnotesize
\TSETabSetup
\caption{Target speaker extraction under sparse overlap on IEMOCAP.
TSE-Omni (V) uses clean visual cues; TSE-Omni (A) uses audio enrollment.
Fine-tuned from the VoxCeleb2 core checkpoint. Best values are bold.}
\label{tab:sparse}
\begin{tabular}{@{} l *{6}{Z} @{}}
\toprule
\multirow{2}{*}{\textbf{Model}} & \multicolumn{3}{c}{\textbf{Semantic-Speaker-Quality$\uparrow$}}
 & \multicolumn{3}{c}{\textbf{DNSMOS$\uparrow$}} \\
\cmidrule(lr){2-4} \cmidrule(lr){5-7}
 & \textbf{SBS} & \textbf{SIM} & \textbf{NISQA} & \textbf{SIG} & \textbf{BAK} & \textbf{OVL} \\
\midrule
USEV~\cite{usev}      & \textbf{0.774} & \textbf{0.872} & 1.300 & 2.620 & 2.600 & 1.980 \\
\textbf{TSE-Omni (V)} & 0.750 & 0.821 & \textbf{1.520} & \textbf{2.850} & \textbf{2.910} & \textbf{2.260} \\
\textbf{TSE-Omni (A)} & 0.743          & 0.817          & 1.510 & 2.840 & 2.870 & 2.240 \\
\bottomrule
\end{tabular}
\end{table}
This trade-off reflects the different behaviors of generative models on
sparse-overlap mixtures: the AR-LLM backbone can leverage visual cues
(e.g., lip non-movement) or pre-enrolled audio cues to predict silent
speech tokens, which occasionally introduces minor false extractions
(lower SBS/SIM) but produces cleaner output on NISQA and DNSMOS. We do
not claim a semantic win on this set.

\subsubsection{Streaming Inference}

MeMo~\cite{li2025memo} is an online AV-TSE framework tailored for
streaming inference. We compare it with TSE-Omni using a chunk size of 1 second,
with results in Table~\ref{tab:streaming_performance}.

\begin{table*}[h]
\centering
\footnotesize
\TSETabSetup
\caption{Streaming on the VoxCeleb2 test set, 1\,s chunks. Qwen2.5 is the 77M
from-scratch backbone (same size as LauraGPT), not Qwen2.5-0.5B. Best values
are bold.}
\label{tab:streaming_performance}
\begin{tabular}{@{} lll *{7}{Y} @{}}
\toprule
\textbf{Model} & \textbf{Backbone} & \textbf{Cue}
 & {\textbf{SBS}} & {\textbf{SIM}} & {\textbf{NISQA}}
 & {\textbf{SIG}} & {\textbf{BAK}} & {\textbf{OVL}} & {\textbf{RTF}$\downarrow$} \\
\midrule
MeMo~\cite{li2025memo} & -- & V & \textbf{0.76} & \textbf{0.95} & 2.02 & 3.15 & 2.71 & 2.35 & \textbf{0.03} \\
\midrule
\multirow{4}{*}{{TSE-Omni}}
 & \multirow{2}{*}{LauraGPT}
   & A & 0.75 & 0.90 & \textbf{2.73} & 3.36 & \textbf{3.21} & \textbf{2.73} & 0.64 \\
   & & V & 0.72 & 0.89 & 2.64 & 3.35 & 3.19 & 2.71 & 0.65 \\
 & \multirow{2}{*}{Qwen2.5}
   & A & 0.64 & 0.73 & 2.53 & \textbf{3.40} & 3.09 & 2.68 & 0.23 \\
   & & V & 0.65 & 0.72 & 2.56 & 3.38 & 3.14 & 2.69 & 0.24 \\
\bottomrule
\end{tabular}
\end{table*}

MeMo achieves slightly better SBS and SIM, while TSE-Omni delivers
superior speech quality across all metrics (e.g., NISQA 2.73 vs.\ 2.02 with
the LauraGPT backbone). Regarding efficiency, MeMo attains the lowest RTF
of 0.03 thanks to its RNN-based computation, but it requires a 2-second
warm-up via a pseudo-AR sliding window before extraction can begin. TSE-Omni
has no such pseudo-AR warm-up; it still uses the first 2\,s of the visual cue
for recognition, as in offline visual TSE. Its RTF is 0.64
with the LauraGPT backbone (limited by autoregressive sampling and
multi-head attention) and reduces to 0.23 with the 77M Qwen2.5 backbone thanks to
grouped-query attention. This latency reduction, however, comes at the
cost of a visible semantic-quality drop (SBS 0.75$\to$0.64 for the audio
cue), a trade-off that should be weighed for each deployment scenario.

The performance drop relative to offline inference is likely due to the
NAR stage, where the vocoder benefits from a broader context
than a single 1-second chunk provides.

\subsection{Potential of Modality Extension and Combination}
\begin{figure}[htbp]
\centering
\includegraphics[width=1.2\linewidth]{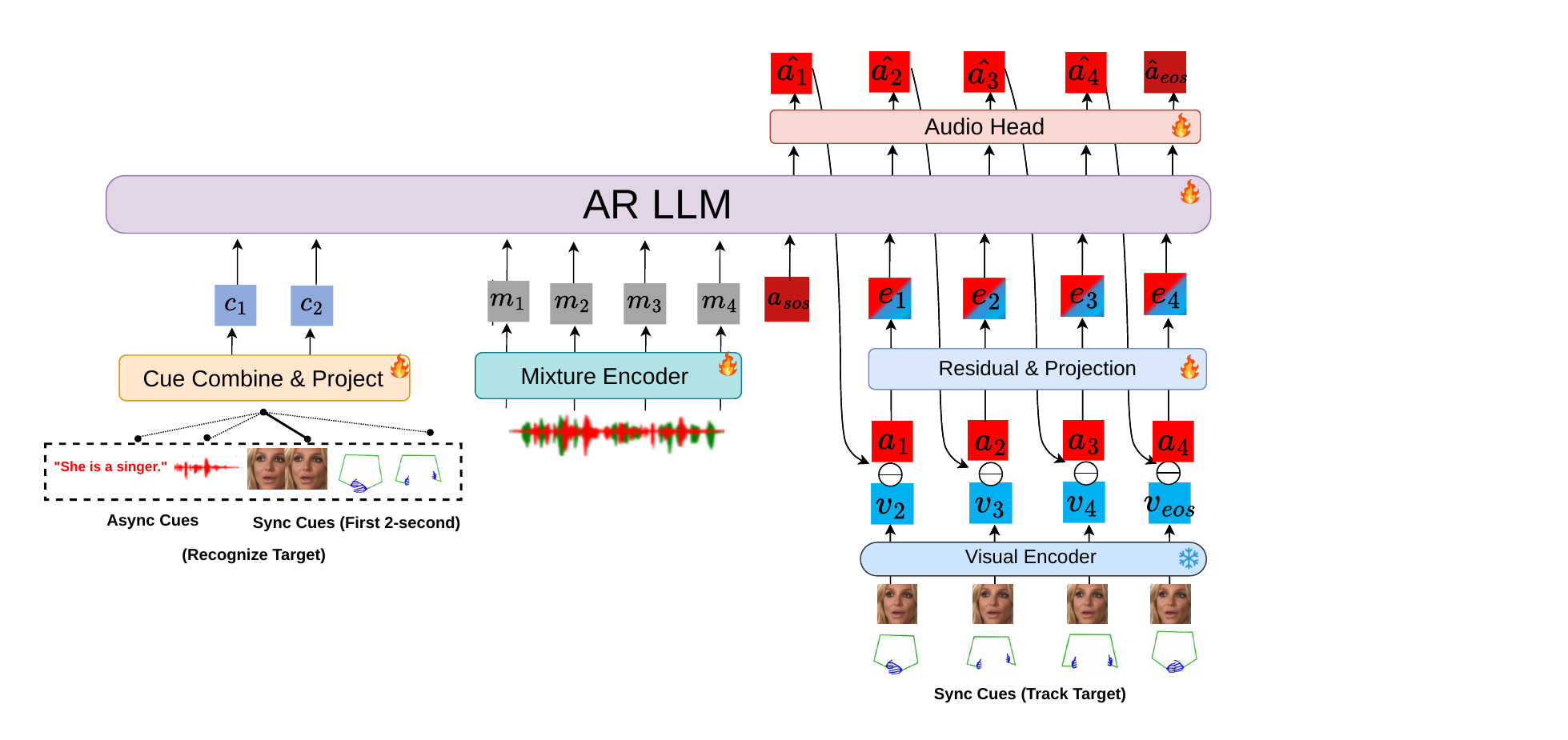}
\caption{Multi-cue fusion: each encoder is mapped by a linear layer into the
shared $D$-dimensional space; a missing cue is a zero tensor. Used for
Table~\ref{tab:modality_all}.}
\label{fig:multi_cues}
\end{figure}

Fig.~\ref{fig:multi_cues} shows how cue embeddings are concatenated. Table~\ref{tab:modality_all} reports Audio, Text, and Visual combinations on VoxCeleb2 and YGD.

\begin{table}[h]
\centering
\footnotesize
\TSETabSetup
\begin{threeparttable}
\caption{Ablation of cue combinations. VoxCeleb2 columns: trained from scratch
on VoxCeleb2 (V = lip movements). YGD columns: trained from scratch on
YGD-2mix (V = co-speech gestures). Not the same checkpoint. Bold = best
among TSE-Omni rows.}
\label{tab:modality_all}
\begin{tabular}{@{} ll *{6}{Z} @{}}
\toprule
\multirow{2}{*}{\textbf{Model}} & \multirow{2}{*}{\textbf{Cue}}
 & \multicolumn{3}{c}{\textbf{Vox2}}
 & \multicolumn{3}{c}{\textbf{YGD}} \\
\cmidrule(lr){3-5} \cmidrule(lr){6-8}
 & {\textbf{SBS}} & {\textbf{SIM}} & {\textbf{OVL}} & {\textbf{SBS}} & {\textbf{SIM}} & {\textbf{OVL}} \\
\midrule
USEV / SEG$^{\dagger}$ & V & 0.800  & 0.940 & 2.600 & 0.650 & 0.840 & 2.580 \\
\midrule
\multirow{7}{*}{\textbf{TSE-Omni}}
 & A     & 0.790 & 0.921 & 2.851 & \textbf{0.713} & 0.835 & 2.893 \\
 & T     & 0.602 & 0.746 & \textbf{2.874} & 0.442 & 0.519 & 2.643 \\
 & V     & 0.803 & 0.937 & 2.831 & 0.602 & 0.732 & 2.776 \\
 & A+T   & 0.771 & 0.925 & 2.848 & 0.704 & \textbf{0.844} & \textbf{2.964} \\
 & A+V   & \textbf{0.810} & \textbf{0.953} & 2.852 & 0.701 & 0.838 & 2.772 \\
 & T+V   & 0.772 & 0.920 & 2.851 & 0.700 & 0.836 & 2.893 \\
 & A+T+V & 0.771 & 0.920 & 2.851 & 0.709 & 0.833 & 2.837 \\
\bottomrule
\end{tabular}
\begin{tablenotes}[flushleft]
\footnotesize
\item $^{\dagger}$Baseline: the conventional visual method per dataset
(Vox2: USEV; YGD: SEG), using $V$;
$V$ is lip movements on Vox2 and co-speech gestures on YGD.
\end{tablenotes}
\end{threeparttable}
\end{table}

Main findings:
\begin{itemize}
    \item \textbf{Combining Audio and Lip Cues Outperforms Either Alone.} Combining audio and lip cues
    (A + V) gives the highest speaker similarity on Vox2
    ($\mathbf{0.953}$), consistent with video helping to track the target speaker.
    \item \textbf{Text Cues Perform Poorly.} Text-only (T) results are weak
    on SBS/SIM (Vox2: 0.602/0.746; YGD: 0.442/0.519), largely because there
    is not enough training data to properly match textual embeddings, mixture
    embeddings, and speech semantic tokens. (Its relatively high Vox2 OVL of
    2.874 notwithstanding, text-only extraction fails to track the target
    identity.)
    \item \textbf{Mixing Strong and Weak Modalities Can Hurt.} Adding a weak
    modality to a strong setup often lowers overall performance, indicating that
    multi-modal fusion remains non-trivial.
   \item \textbf{Co-speech gesture cues alone are not sufficient.} On YGD,
    gesture cues (V) perform worse than lip movements (Vox2's V) because
    body language mainly provides prosody and lacks the detailed visemes
    found in lip movements (SBS 0.602 vs.\ 0.803). In addition, directly
    upsampling gestures to match the frame rate of speech semantic tokens is
    ineffective.
\end{itemize}

\subsection{Ablation Studies}

\subsubsection{LLM Backbones and Initialization}
We examine three questions: \textbf{Q1}, whether a more advanced LLM
architecture helps (LauraGPT vs.\ Qwen2.5, both trained from scratch);
\textbf{Q2}, whether pretrained textual weights are beneficial (Qwen2.5-0.5B
with vs.\ without initialization); and \textbf{Q3}, whether fine-tuning from
the official LauraTSE checkpoint outperforms training from scratch. Results
are reported in Table~\ref{tab:llm_full}.

\begin{table}[h]
\centering
\footnotesize
\TSETabSetup
\begin{threeparttable}
\caption{LLM architecture, textual initialization, and LauraTSE initialization.
All rows are trained on VoxCeleb2 (in-domain).}
\label{tab:llm_full}
\begin{tabular}{@{} llll *{4}{Z} @{}}
\toprule
\textbf{Backbone} & \textbf{Param} & \textbf{Mode} & \textbf{Cue}
 & {\textbf{SBS}} & {\textbf{SIM}} & {\textbf{NISQA}} & {\textbf{OVL}} \\
\midrule
\multicolumn{8}{@{}l}{\textit{Q1: LLM architecture (from scratch)}} \\
\midrule
\multirow{2}{*}{LauraGPT} & \multirow{2}{*}{77M} & \multirow{2}{*}{Scr} & A & 0.783 & 0.931 & 3.230 & 2.840 \\
 & & & V & \textbf{0.806} & \textbf{0.941} & \textbf{3.240} & \textbf{2.850} \\
\multirow{2}{*}{Qwen2.5} & \multirow{2}{*}{77M} & \multirow{2}{*}{Scr} & A & 0.706 & 0.900 & 3.180 & 2.740 \\
 & & & V & 0.710 & 0.893 & 3.220 & 2.770 \\
\midrule
\multicolumn{8}{@{}l}{\textit{Q2: Pretrained textual weights (Qwen2.5-0.5B)}} \\
\midrule
\multirow{4}{*}{Qwen2.5} & \multirow{4}{*}{0.5B} & \multirow{2}{*}{Scr} & A & 0.489 & 0.550 & 1.873 & 2.510 \\
 & & & V & 0.491 & 0.550 & 1.834 & 2.456 \\
 & & \multirow{2}{*}{Init} & A & 0.667 & 0.728 & 2.786 & 2.857 \\
 & & & V & \textbf{0.692} & \textbf{0.747} & \textbf{2.796} & \textbf{2.859} \\
\midrule
\multicolumn{8}{@{}l}{\textit{Q3: Initialization from LauraTSE}} \\
\midrule
\multirow{2}{*}{LauraGPT} & \multirow{2}{*}{77M} & \multirow{2}{*}{Scr} & A & 0.783 & 0.931 & 3.230 & 2.840 \\
 & & & V & 0.806 & \textbf{0.941} & \textbf{3.240} & 2.850 \\
\multirow{2}{*}{LauraGPT$^{\dagger}$} & \multirow{2}{*}{77M} & \multirow{2}{*}{Init} & A & \textbf{0.807} & 0.900 & 3.090 & 2.849 \\
 & & & V & 0.789 & 0.902 & 3.109 & \textbf{2.868} \\
\bottomrule
\end{tabular}
\begin{tablenotes}[flushleft]
\footnotesize
\item \textbf{Scr}: from scratch; \textbf{Init}: from a pretrained checkpoint
(Qwen2.5-0.5B in Q2, LauraTSE in Q3). A/V: audio/visual cue.
\textbf{Bold}: best per block.
\item $^{\dagger}$ LauraTSE checkpoint (LibriSpeech/Libri2Mix), fine-tuned on VoxCeleb2.
\end{tablenotes}
\end{threeparttable}
\end{table}
Three observations can be drawn from Table~\ref{tab:llm_full}.
\textbf{(1)} With a comparable parameter budget (77M) and training recipe,
the LauraGPT-style backbone consistently outperforms the downscaled
Qwen2.5 (e.g., visual-cue SBS 0.806 vs.\ 0.710). Despite similar
parameters, Qwen2.5 uses a more sophisticated architecture than the compact
LauraGPT decoder; its higher capacity makes it prone to overfitting on the
limited training data, so the simpler, speech-adapted LauraGPT backbone
generalizes better.
\textbf{(2)} Trained from scratch, Qwen2.5-0.5B fails to learn the task
(SBS $<$ 0.50), whereas initializing with pretrained textual weights lifts
SBS to 0.67--0.69, confirming the value of pretrained linguistic knowledge;
nevertheless, it still lags behind the 77M LauraGPT backbone under our data
budget.
\textbf{(3)} Initializing from the LauraTSE checkpoint improves audio-cue
SBS (0.807 vs.\ 0.783) but slightly lowers visual-cue SBS and NISQA, which
we attribute to the domain gap between LibriSpeech/Libri2mix pretraining
and VoxCeleb2 fine-tuning. On Libri2mix, where baselines are in-domain,
TSE-Omni (A) is zero-shot (SBS 0.83). TSE-Omni (A+Pretrain) is LauraTSE
initialization plus VoxCeleb2 fine-tuning and reaches SBS 0.88
(Table~\ref{tab:unified_compact_full}); that row should not be read as a
Libri2mix-trained TSE-Omni.

\subsubsection{Audio-Visual Token Integration Strategy}

\begin{figure}[htbp]
\centering
\includegraphics[width=0.98\linewidth]{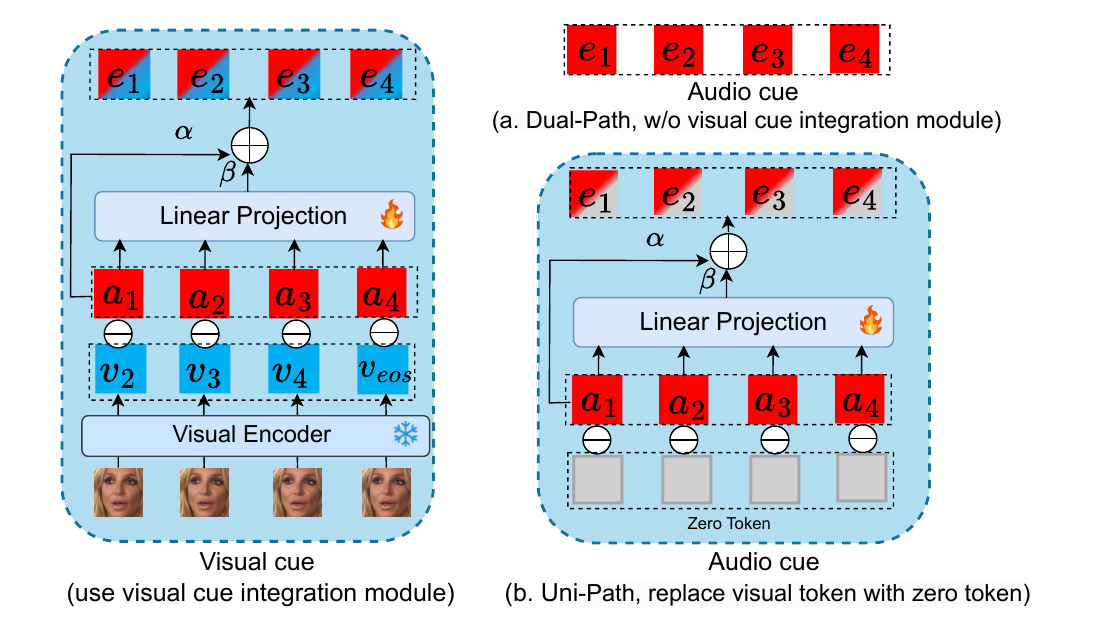}
\caption{Token integration. Dual-Path: the visual module is used only with a
visual cue. Uni-Path: one module for both cues, with zeros in place of visual
tokens for audio-cue TSE.}
\label{fig:dual_single_path}
\end{figure}

\noindent\textbf{Dual-Path vs.\ Uni-Path.}
We compare two token integration strategies, illustrated in
Fig.~\ref{fig:dual_single_path}; results are reported in
Table~\ref{tab:cue_strategy_comparison}.

\begin{table}[h]
\centering
\footnotesize
\TSETabSetup
\caption{Dual-Path vs.\ Uni-Path token integration on the LRS3 two-speaker
test set (zero-shot from the VoxCeleb2 core checkpoint).
Best values within each cue are bold.}
\label{tab:cue_strategy_comparison}
\begin{tabular}{@{} ll *{6}{Y} @{}}
\toprule
\multirow{2}{*}{\textbf{Cue}} & \multirow{2}{*}{\textbf{Strategy}}
 & \multicolumn{3}{c}{\textbf{Semantic-Speaker-Quality$\uparrow$}}
 & \multicolumn{3}{c}{\textbf{DNSMOS$\uparrow$}} \\
\cmidrule(lr){3-5} \cmidrule(lr){6-8}
 & & \textbf{SBS} & \textbf{SIM} & \textbf{NISQA} & \textbf{SIG} & \textbf{BAK} & \textbf{OVL} \\
\midrule
\multirow{2}{*}{Audio} & Dual-Path & \textbf{0.91} & \textbf{0.97} & \textbf{3.89} & \textbf{3.50} & \textbf{3.79} & \textbf{3.08} \\
 & Uni-Path  & 0.89 & \textbf{0.97} & 3.72 & 3.46 & 3.73 & 3.05 \\
\midrule
\multirow{2}{*}{Visual} & Dual-Path & \textbf{0.89} & \textbf{0.96} & \textbf{3.84} & \textbf{3.51} & \textbf{3.76} & \textbf{3.08} \\
 & Uni-Path  & 0.88 & 0.95 & 3.79 & 3.48 & 3.75 & 3.05 \\
\bottomrule
\end{tabular}
\end{table}

Dual-Path consistently outperforms Uni-Path for both cues, with the largest
gap in speech quality (NISQA 3.89 vs.\ 3.72 for audio cues; 3.84 vs.\ 3.79
for visual cues). Uni-Path forces audio-cue extraction to pass through the
visual integration module with zero-padded visual tokens, introducing an
unnecessary distribution shift. Dual-Path also avoids feeding zeros during
\emph{clean} audio-cue inference. Visual-corruption inference is different:
later visual frames are zeros, but the visual module stays on, which is the
protocol of Sec.~\ref{sec:self-enroll}. We adopt Dual-Path
and enable the visual integration module only when a visual cue is used.

\noindent\textbf{Residual Scales.}
We further investigate the gating scales in the visual integration module,
where $\alpha$ weights the historical speech semantic tokens and $\beta$
weights the fused audio-visual tokens; training and inference use
consistent settings. Results are reported in
Table~\ref{tab:residual_scale_ablation}.

\begin{table}[h]
\centering
\footnotesize
\TSETabSetup
\caption{Effect of the residual scales on the LRS3 two-speaker test set
(zero-shot from the VoxCeleb2 core checkpoint).
$\alpha$: scale of the speech semantic tokens; $\beta$: scale of the fused
audio-visual tokens. Best values are bold.}
\label{tab:residual_scale_ablation}
\begin{tabular}{@{} cc *{6}{Y} @{}}
\toprule
\multirow{2}{*}{$\alpha$} & \multirow{2}{*}{$\beta$}
 & \multicolumn{3}{c}{\textbf{Semantic-Speaker-Quality$\uparrow$}}
 & \multicolumn{3}{c}{\textbf{DNSMOS$\uparrow$}} \\
\cmidrule(lr){3-5} \cmidrule(lr){6-8}
 & & \textbf{SBS} & \textbf{SIM} & \textbf{NISQA} & \textbf{SIG} & \textbf{BAK} & \textbf{OVL} \\
\midrule
0  & 1  & 0.79 & 0.92 & 3.74 & 3.33 & 3.74 & 2.94 \\
1  & 1  & \textbf{0.89} & \textbf{0.96} & \textbf{3.84} & \textbf{3.51} & \textbf{3.76} & \textbf{3.08} \\
1  & 10 & 0.86 & 0.96 & 3.81 & 3.46 & 3.72 & 3.01 \\
10 & 1  & 0.88 & 0.96 & 3.64 & 3.43 & 3.61 & 2.95 \\
\bottomrule
\end{tabular}
\end{table}

Equal scales ($\alpha=\beta=1$) give the best performance across all
metrics. Setting $\alpha=0$, i.e., relying solely on fused audio-visual
tokens without the speech residual, drops SBS to 0.79, confirming that the
speech-semantic stream remains the primary information source for
extraction. Conversely, over-weighting the self-enrollment branch
($\alpha=10$) causes the largest quality drop (NISQA 3.64), suggesting that
exclusively trusting historical predictions amplifies error accumulation.
Overall, the residual connection from historical speech tokens is
essential: it is this connection that keeps self-enrollment available when
later visual frames are zeros. It still has to be balanced with the visual
branch.

\subsubsection{Stronger Visual Semantic Tokens}
\label{sec:visual_tokens}
We explore two ways to strengthen the visual representation: replacing the
lightweight VSR (ResNet) encoder with the heavier AV-HuBERT (Large) as the
visual encoder, and adding an auxiliary visual head that predicts discrete
visual semantic tokens. As shown in
Table~\ref{tab:visual_strategy_ablation}, neither change improves
extraction. The stronger encoder yields an SBS virtually identical to
ResNet's, which is consistent with the limited capacity of the 77M
backbone to absorb deeper visual representations; adding the visual head
slightly lowers SBS. We therefore keep the lightweight ResNet and \emph{omit}
the visual head in the default model (Table~\ref{tab:unified_compact_full}).
The auxiliary visual loss is reported only here; we do not use it to claim
extra viseme--phoneme regularization in the main system.

\begin{table}[h]
\centering
\footnotesize
\TSETabSetup
\caption{Visual modeling strategies on the LRS3 two-speaker test set
(zero-shot from the VoxCeleb2 core checkpoint).
\textbf{In-V~Rep.}: input visual representation. \textbf{V-Loss}: whether
the visual CE loss (and hence the visual head) is enabled.
$\Delta$\textbf{SBS} is the change in SBS relative to the best value in the
table ($0.895$). All configurations are within $\pm 0.011$ of this best,
i.e., differences are statistically negligible. The default model uses
VSR (ResNet) without V-Loss.}
\label{tab:visual_strategy_ablation}
\begin{tabular}{@{} ll Z S[table-format=-1.3] @{}}
\toprule
\textbf{In-V Rep.} & \textbf{V-Loss}
 & \textbf{SBS} & {$\Delta$SBS} \\
\midrule
VSR (ResNet)  & $\times$     & 0.894 & -0.001 \\
VSR (ResNet)  & $\checkmark$ & 0.887 & -0.008 \\
AV-HuBERT (L) & $\times$     & \textbf{0.895} & 0.000 \\
AV-HuBERT (L) & $\checkmark$ & 0.884 & -0.011 \\
\bottomrule
\end{tabular}
\end{table}

\subsubsection{Cue Robustness}
We further study the cue injection mechanism and answer one question: is it
beneficial to enroll the cue at both the AR and NAR stages, or is the AR
stage sufficient? Results are summarized in
Table~\ref{tab:enroll_stage_ablation}.

\begin{table}[h]
\centering
\footnotesize
\TSETabSetup
\caption{Effect of the cue enrollment stage on the LRS3 two-speaker test set
(zero-shot from the VoxCeleb2 core checkpoint).
Best values within each cue are bold.}
\label{tab:enroll_stage_ablation}
\begin{tabular}{@{} ll *{6}{Z} @{}}
\toprule
\multirow{2}{*}{\textbf{Cue}} & \multirow{2}{*}{\textbf{Enr.~Stage}}
 & \multicolumn{3}{c}{\textbf{Semantic-Speaker-Quality$\uparrow$}}
 & \multicolumn{3}{c}{\textbf{DNSMOS$\uparrow$}} \\
\cmidrule(lr){3-5} \cmidrule(lr){6-8}
 & & \textbf{SBS} & \textbf{SIM} & \textbf{NISQA} & \textbf{SIG} & \textbf{BAK} & \textbf{OVL} \\
\midrule
\multirow{2}{*}{Audio}
 & AR     & 0.894 & 0.967 & 3.680 & 3.460 & 3.740 & 3.020 \\
 & AR+NAR & \textbf{0.914} & \textbf{0.973} & \textbf{3.890} & \textbf{3.500} & \textbf{3.790} & \textbf{3.080} \\
\midrule
\multirow{2}{*}{Visual}
 & AR     & 0.886 & 0.960 & 3.820 & 3.490 & 3.740 & 3.050 \\
 & AR+NAR & \textbf{0.894} & \textbf{0.960} & \textbf{3.840} & \textbf{3.510} & \textbf{3.760} & \textbf{3.080} \\
\bottomrule
\end{tabular}
\end{table}

\noindent\textbf{Enrollment stage.}
Enrolling the cue at both stages consistently outperforms AR-only
enrollment across all metrics for both cues. The NAR Conformer and codec
decoder are trained with this extra cue input; the gain is therefore not from a
frozen vocoder that has never seen the cue.

\section{Conclusion}
We have presented TSE-Omni, a unified framework that performs both audio- and
visual-cue target speech extraction with a single autoregressive LLM,
removing per-modality models and reducing deployment cost. TSE-Omni predicts
target speech semantic tokens from its own prediction history
(self-enrollment), fused with synchronized visual cues via a residual module.
This yields cross-modality compensation without corruption-specific training:
when visual cues are degraded or missing, extraction continues from the
accumulated speech-semantic context, with no visual-recovery module. With 
few parameters, TSE-Omni matches or surpasses state-of-the-art baselines
across most scenarios and clearly leads under visual corruption, speaker
switching, and sparse overlap, while supporting streaming inference and
extending to further cues such as text and co-speech gestures. Future work
will explore stronger LLM backbones, more cue modalities, and lower streaming
latency while improving robustness.

\bibliographystyle{IEEEtran}
\bibliography{refs}

\end{document}